\documentclass[useAMS, usenatbib]{mn2e}

\usepackage{aas_macros}
\usepackage{graphicx}  
\usepackage{multirow}

\title[The IR/X-ray correlation of GX 339-4 ]{The infrared/X-ray correlation of GX 339-4: Probing hard X-ray emission in accreting black holes.}
\author[M. Coriat et al.]{M. Coriat,$^{1}$\thanks{E-mail: mickael.coriat@cea.fr}
S. Corbel,$^{1}$ M. M. Buxton,$^{2}$ C. D. Bailyn,$^{2}$ 
\newauthor J. A. Tomsick,$^{3}$ E. K\"{o}rding$^{1}$ and E. Kalemci$^{4}$\\
$^{1}$Universit\'e Paris Diderot and Service d'Astrophysique, UMR AIM, CEA Saclay, F-91191 Gif-sur-Yvette, France.\\
$^{2}$ Astronomy Department, Yale University, P.O. Box 208101, New Haven, CT 06520-8101, USA.\\
$^{3}$ Space Sciences Laboratory, 7 Gauss Way, University of California, Berkeley, CA94720-7450, USA.\\
$^{4}$ Sabanc\i\ University,Orhanl\i-Tuzla 34956, \.Istanbul, Turkey.}
\begin{document}

\date{\today }

\pagerange{\pageref{firstpage}--\pageref{lastpage}} \pubyear{2009}

\maketitle

\label{firstpage}

\begin{abstract}
GX 339-4 has been one of the key sources for unravelling the accretion ejection coupling in accreting
stellar mass black holes. After a long period of quiescence between 1999 and 2002, GX 339-4
underwent a series of 4 outbursts that have been intensively observed by many ground based
observatories [radio, infrared(IR), optical] and satellites (X-rays). Here, we present results of these broad-band observational campaigns, focusing on the optical-IR (OIR)/X-ray 
flux correlations over the four outbursts. We found tight OIR/X-ray 
correlations over four decades with the presence of a break in the IR/X-ray correlation in the hard state. This correlation is the same for all four outbursts.
This can be interpreted in a consistent way by considering a synchrotron self-Compton origin of the X-rays
 in which the break frequency varies between the optically thick and thin regime of the jet spectrum. We also highlight the similarities and differences between optical/X-ray and 
 IR/X-ray correlations which suggest a jet origin of the near-IR emission in the hard state while the optical is more likely dominated by the blackbody emission of the 
 accretion disc in both hard and soft state. However we find a non negligible contribution of 40 per cent of the jet emission in the \textit{V} band during the hard state.
 We finally concentrate on a soft-to-hard state transition during the decay of the 2004 outburst by comparing the radio, IR, optical and hard X-rays light curves. 
 It appears that unusual delays between the peak of emission in the different energy domains may provide some important constraints on jet formation scenario.
 
 \end{abstract}


\section{Introduction}

Black hole X-ray binaries (BHXBs) are known to be powerful multi wavelength emitters spending most of their life in quiescence
and undergoing sporadic outbursts during which their X-ray luminosities can increase up to a factor 10$^6$ compared to quiescent level. 
During these outbursts, BHXBs display a sequence of  three main X-ray states,  which are defined 
by their X-ray spectral and timing properties.  The hard state, is usually observed at the beginning and 
the end of a `typical' outburst. The X-ray spectra are dominated by  non-thermal emission in the form of a power-law
extending up to hard X-rays. The physical origin of this emission is still debated, but the common models imply a corona of hot plasma
surrounding the compact object or the base of compact jets (\citealt{zdziarski98}; \citealt*{nowak02,markoff05}). 
The soft state is dominated by thermal emission from the accretion disc and
 displays a weak power-law component with photon index softer than during the hard state. The transitional states between the hard state and the soft state can be classified 
 as various instances (hard or soft) of the Intermediate state and has spectra with hardnesses in between the hard and the soft states.
 For a complete description of X-ray states properties and classification see e.g. \citet{homan05a},  \citet{mcclintock06} and \citet{fender06}.
 
Among the canonical states, it is perhaps the hard state that has attracted most  attention in recent years where radio observations 
 have highlighted analogies with the flat spectra of low-luminosity active galactic nuclei \citep[AGN; ][]{falcke96a}.
The radio spectrum from GX 339--4 during its hard state can be explained as self-absorbed synchrotron emission from compact jets 
\citep{corbel00, fender01}, similar to the case of flat spectrum AGN \citep{blandford79}. Regular radio observations 
have shown that the compact jets of GX 339--4 are quenched in the soft state \citep{fender99,corbel00}.  Similar properties have 
now been found in a growing number of BHXBs (the jet has even been resolved in Cyg X--1, \citealt{stirling01}, and GRS 1915+105, \citealt*{dhawan00}; \citealt{fuchs03}), thus suggesting
that compact jets are ubiquitous in BHXBs during the hard state \citep{fender01}.  In addition to being responsible for the radio emission, 
the compact jets may dominate the infrared emission and could also contribute significantly in the optical domain \citep{corbel01,jain01,corbel02,
homan05,russell06}. The compact jets may also produce some of the observed X-ray
emission \citep{markoff03,markoff05}. Thus, the compact jets may dominate large fractions of the spectral energy distribution (SED) of BHXBs  (\citealt{fender01,corbel02}; \citealt*{gallo03}).

A good way to assess the contribution of jets at high energy and disentangle its emission components, is to perform broad-band
observations of  BHXB simultaneously at radio, optical-infrared (OIR) and X-ray frequencies, and study the correlations between 
these frequency domains. Radio/X-ray  \citep{corbel00,corbel03,gallo03} and OIR/X-ray  \citep{homan05,russell06} flux correlations 
have already been found for BHXBs in their hard state indicating a strong coupling between accretion and ejection processes. 
GX 339-4 was the first galactic black hole for which a strong non-linear correlation between radio and X-ray emission was observed
in the hard state \citep{corbel00,corbel03}.  This correlation has been extended to other galactic black holes \citep{gallo03} and even AGN \citep*{merloni03,falcke04,kording06a}.
In this work, we present the detailed evolution of the IR/X-ray flux correlation of  GX 339-4 over its last five years of activity. 
In section 2, we describe the observational data set and our analysis method. 
After a brief description of the recent activity of GX 339-4, section 3 presents the results of our broad-band study and  we discuss and propose
 possible interpretations in section 4. Our conclusions are summarised in section 5.

\section[]{Observations and data analysis}

\subsection{Optical and Infrared}

Optical and IR photometry of GX 339--4 were conducted between UT 2002 January 22
and 2007 October 02 (MJD 52296--54375). Approximately 840 observations were made with the SMARTS (Small and Medium Aperture Research
Telescope System) which currently uses ANDICAM camera on the 1.3m CTIO telescope.
ANDICAM takes simultaneous optical and IR images over a variety of band passes. In the observations reported here, we used Johnson-Kron-Cousins $V$ and
$I$ filters \citep*{bessell98} and standards CIT/CTIO $J$ and $H$ filters \citep{elias82}. The optical and IR light curves for the all period are presented in \citet{buxton07}.
We converted the observed magnitudes $m$ into spectral flux densities $F_{\nu}$ using
 the optical extinction $A_{V} = 3.7 \pm 0.2$ \citep{zdziarski98} and the extinction law of 
\citet*{cardelli89}. The uncertainty on the optical extinction 
has been propagated in the derived errors on the OIR fluxes and dominates over the intrinsic errors.

\subsection{X-ray}

\subsubsection{Data reduction}

The X-ray observations were performed  with the Proportional Counter
Array (PCA) and the High Energy X-ray Timing Experiment (HEXTE) on-board
the {\it Rossi X-ray Timing Explorer} (RXTE). We analysed public data taken between 2002 January
29 (MJD 52303) and 2007 October 6 (MJD 54379), corresponding to
a total of 622 pointed observations.  Spectra were produced for each
observation using  \textsc{heasoft V6.4}.

For the PCA, only data from Proportional Counter Unit 2
(PCU2) were used for the analysis in this work, as it is the only operational
unit during all observations and is the best calibrated detector out of the 5 PCUs.
PCA spectra were extracted from the \textsc{standard 2} mode data and a systematic 
error of 0.6 per cent was added.

For the HEXTE, starting on 2005 December 12, we only used cluster B data
as cluster A started to lose its background measurements capabilities. 
HEXTE spectra were produced from the standard mode \textsc{archive} data
and were dead-time corrected.

\subsubsection{Spectral analysis and state classification}

The  PCA (3--25 keV) and HEXTE (20--150 keV) spectra of each
observation were fitted simultaneously in {\tt XSPEC V11.3.2}
using an overall normalization constant that was
allowed to float for cross-calibration purposes. To fit the spectra, we used 
several combinations of the following models: a (cut-off) power-law ({\tt
cutoffpl} or {\tt powerlaw}), a multi-temperature disc blackbody model ({\tt ezdiskbb}), a
Gaussian emission line at 6.4 keV ({\tt gaussian}), a
smeared absorption edge ({\tt smedge}), and an absorption component
({\tt wabs}). The hydrogen column density, $N_H$, was fixed to a value of
$5\times10^{21}$  atoms cm$^{-2}$ \citep{kong00}.
At fainter flux, when GX 339-4 was not significantly detected with HEXTE, fits
were made to the PCA spectrum only. 
We finally obtained an average reduced $\chi^2$ of 0.96 with a minimum of 0.35 and a maximum of 1.87.
Unabsorbed fluxes were measured in the 3--9, 9--20 and 20--100 keV energy ranges (if HEXTE data were used). 

We classified the observations into the various X-ray states (hard state, intermediate state, soft state) using the dates of the state transitions 
provided by previous spectral and timing studies (\citealt{belloni05a}; \citealt*{smith05}; \citealt{belloni06,kalemci07,del-santo09}). When not available,
we determined the state using the value of the power-law photon index (hard state: $\Gamma<2.1$, soft state: $\Gamma>2.1$) 
and the evolution of the hardness ratio.

\subsubsection{Galactic ridge emission}

At very  low flux level ($\la 10^{-11} \rm \, erg \,s^{-1} cm^{-2}$), X-ray emission from the Galactic ridge \citep{valinia98} 
can significantly contaminate the estimated GX 339-4  fluxes. In order to subtract this contribution, we used quasi-simultaneous
 {\it Chandra} and \textit{RXTE}/PCA observations of GX 339-4 taken on MJD 52911 during a quiescent state of GX 339-4 \citep*{gallo03a}, to 
 estimate the Galactic ridge flux, $F_{GR}$. We analysed the seven PCA observations over this period 
 and combined the spectra to extract an average 3--9 keV flux. The high spatial  resolution of {\it Chandra} provides an estimated real flux for GX 339-4
 and is then subtracted from the averaged PCA flux.  We finally obtained, in the 3--9 keV range, 
 $F_{GR} = (4.2 \pm 0.5) \times 10^{-12} \rm \, erg \,s^{-1} cm^{-2}$. Consequently, we subtracted this value to all 3--9 keV fluxes.

 \subsection{Radio}
Since 1996, we have been performing regular radio observations with the Australia
Telescope Compact Array (ATCA) several times a year and during each outburst (e.g.  
\citealt{fender97, corbel00, corbel03}).
In this paper, we focus on the radio data up to 2005 that is 
presented in Corbel et al. (in preparation). 
The ATCA synthesis telescope is an east-west array consisting of six 22 m
antennas with baselines ranging from 31 m to 6 km.  The continuum
observations have been mainly performed in two frequency bands (with a
total bandwidth of 128 MHz for 32 channels), usually at 4800 MHz 
and 8640 MHz  simultaneously. All ATCA results are
summarised in Corbel et al. (in preparation), which also includes further information on the
radio analysis as well as details on the evolution of the radio/X-ray correlation of GX 339-4. 

\begin{figure*}
    \includegraphics[width=1.0\textwidth]{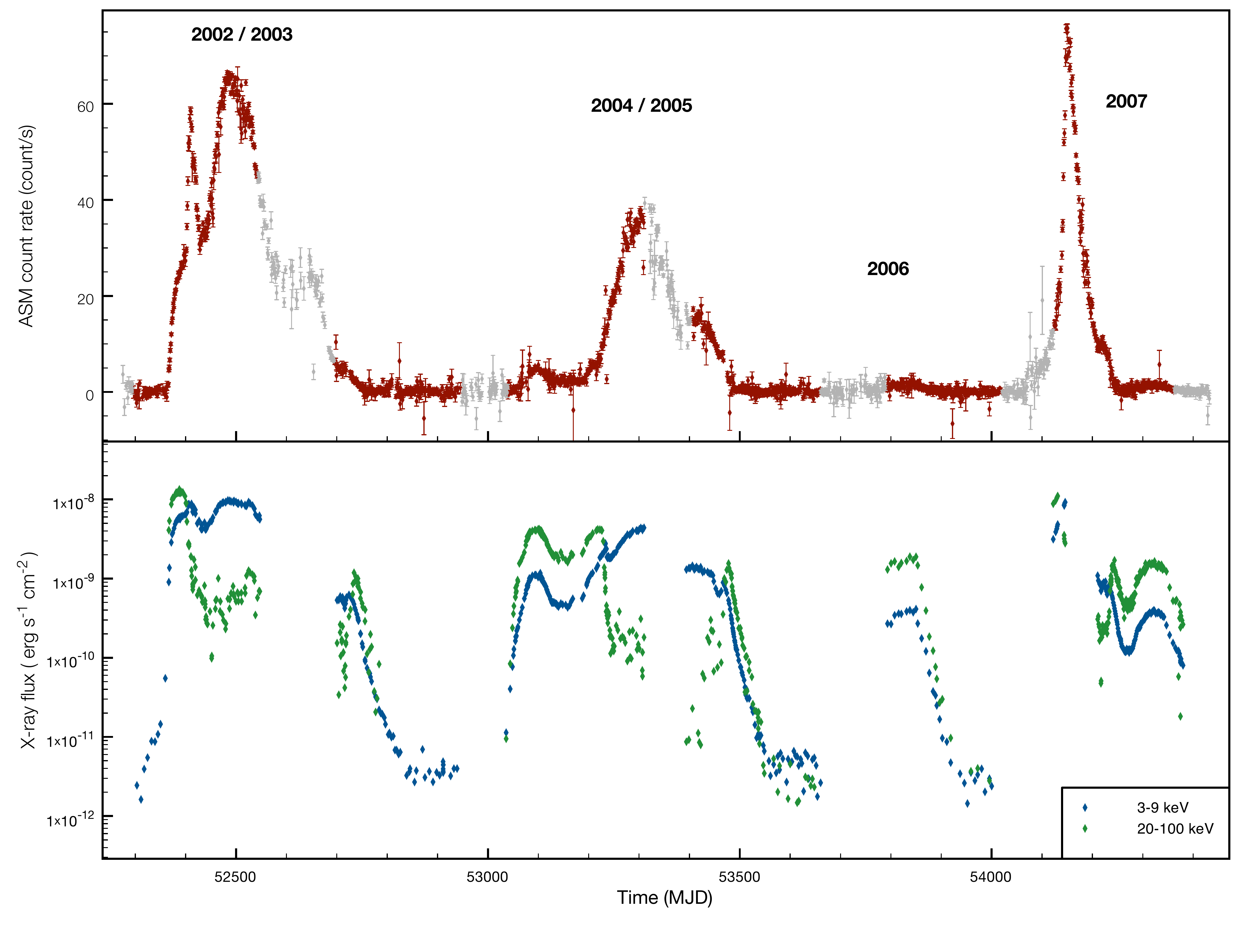} 
          \caption{\textit{Upper panel}: RXTE/ASM light curve of the four GX 339-4 outbursts between 2002 and 2007. The red parts of the light curve represent the periods 
          covered simultaneously in X-ray and OIR. \textit{Lower panel}: RXTE/PCA and RXTE/HEXTE X-ray light curves in the 3--9 keV and 20--100 keV ranges corresponding to the set of observations analysed in this work.}
  \label{asm}
  \end{figure*}

\section[]{Results}

\subsection{Five years of activity}

Fig. \ref{asm} represents the \textit{RXTE}/ASM light curve between 2002 and 2007 (upper panel) and the light curves in the 3--9 keV and 20--100 keV ranges for the 622 analysed RXTE observations (lower panel). During this period, GX 339-4 underwent a series of four outbursts separated by periods of quiescence. The first one and one of the most luminous outburst of those recently observed for this source began in 2002 and ended in 2003 \citep{miller04,belloni05a,homan05}. During the 2002 hard-to-soft state transition, a bright radio flare was observed \citep{fender02}, which later on lead  to the formation of a large-scale relativistic jet \citep{gallo04}. GX 339-4 entered a second outburst in 2004 and went back to quiescence in 2005 \citep{belloni06,miller06,joinet07}. During this outburst, Corbel et al. (in preparation) show the presence of two parallels tracks in the radio/X-ray correlation between the rising and decaying phases of the outburst. During the following `minor' outburst in 2006, GX 339-4 remained in the canonical hard state and reached, in the 3--100 keV band, a maximum unabsorbed flux of $2.75\times 10^{-9} \rm \, erg \,s^{-1} \,cm^{-2}$ (See Fig. \ref{asm}). The last outburst occurred in 2007 \citep{tomsick08,del-santo09} and displayed a luminosity similar to the one reached in 2002. During the initial hard state of this outburst, \citet{miller08} reported the detection of relativistically broadened iron emission line in the X-ray spectra which would suggest a black hole spin parameter close the maximal value. In the following, we will refer to these outbursts by their starting year, namely, 2002, 2004, 2006 and 2007.

\subsection{OIR/X-ray correlations}

Previous studies of several BHXBs in the hard state  \citep{jain01,corbel01,corbel02,buxton04,homan05,russell06}, provide a basic picture of SED, 
with the compact jets emission dominating from radio up to near-IR (NIR), the accretion disc dominating the optical-to-UV range and a power law component
(corona or base of the jets) dominating the soft-to-hard X-rays. 

\begin{figure*}
\centering
\begin{minipage}{1.0\textwidth}
    \includegraphics[width=0.5\textwidth]{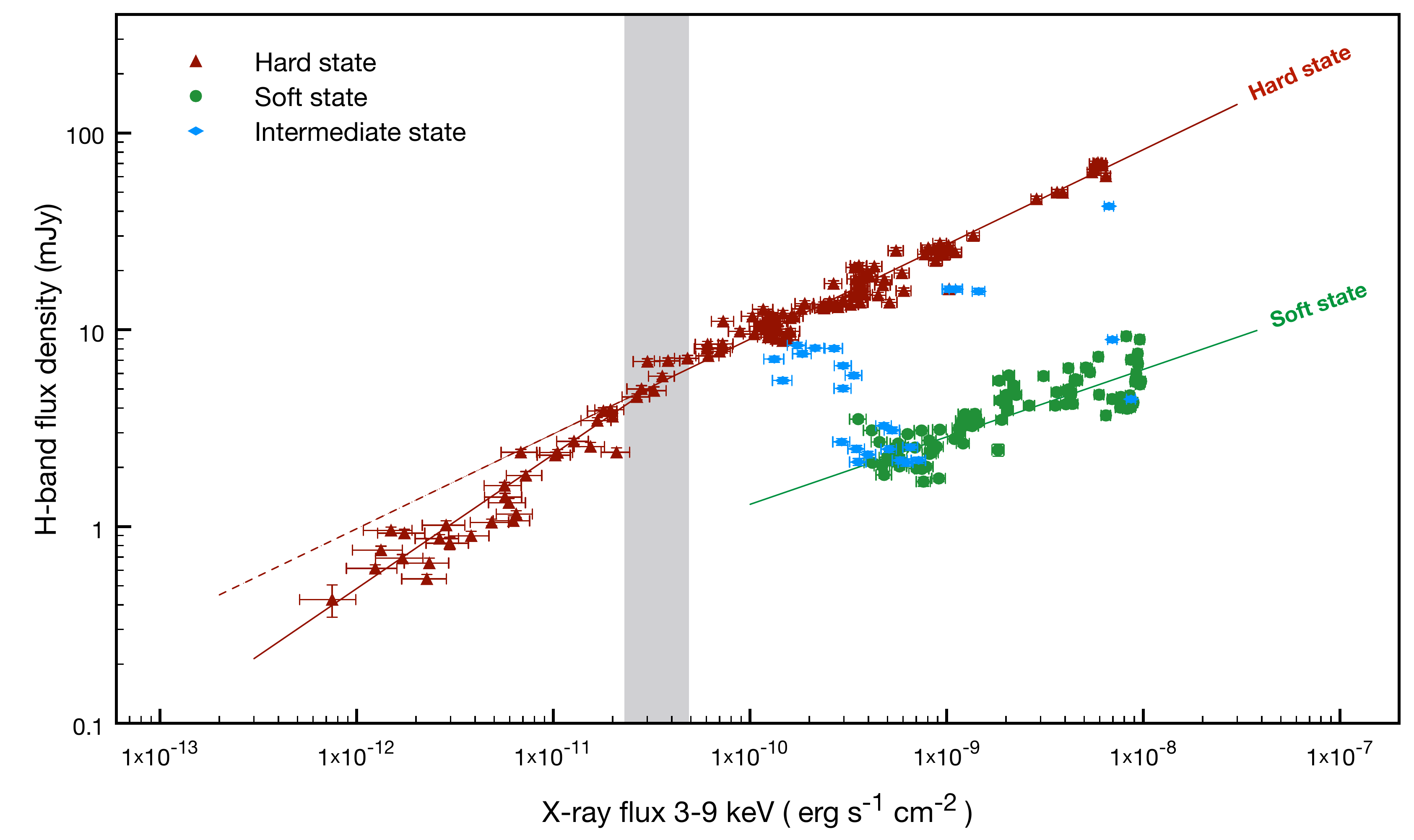} 
        \includegraphics[width=0.5\textwidth]{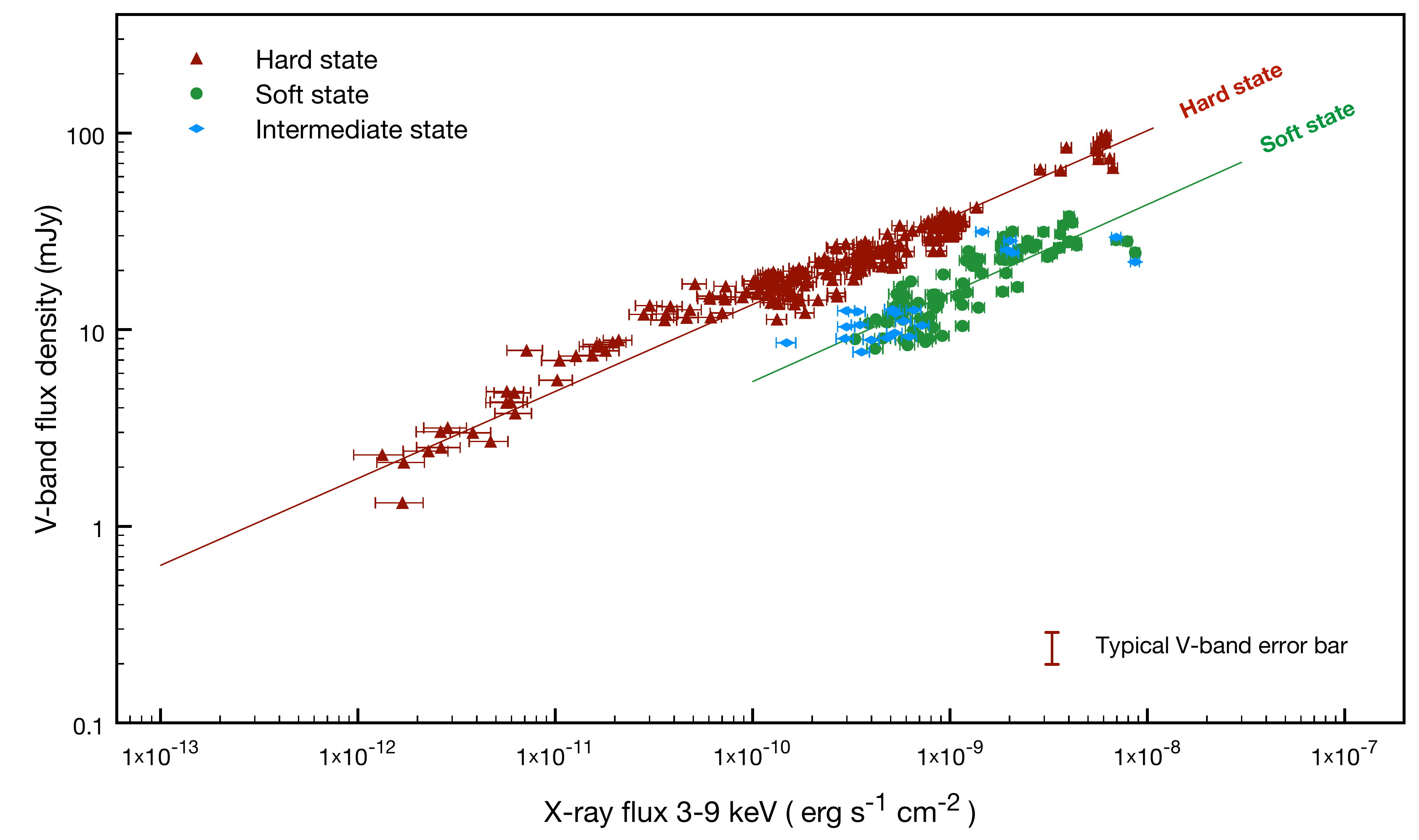} 
           \caption{\textit{(a) Left panel:} Quasi-simultaneous infrared \textit{H-}band flux density versus  3-9 keV X-ray flux during the four outbursts. Data are separated by X-ray states. Red, green and blue points correspond to hard, soft and intermediate state respectively. Red and green continuous lines indicate the fit to the hard state and the soft  state respectively. Grey zone indicates the 90 per cent confidence interval of the X-ray break flux $F_{\rm break}$. \textit{(b) Right panel:} Quasi-simultaneous optical \textit{V-}band flux density versus 3-9 keV X-ray flux during the four outbursts. Data are separated by X-ray states. Red, green and blue data correspond to hard, soft and intermediate state respectively. Red and green continuous lines indicate hard state fit and soft state fit respectively. Both plots have been made with the same scales for comparison purposes.}
  \label{HVXhs}
              \end{minipage}
  \end{figure*}

\begin{figure*}
    \includegraphics[width=160mm]{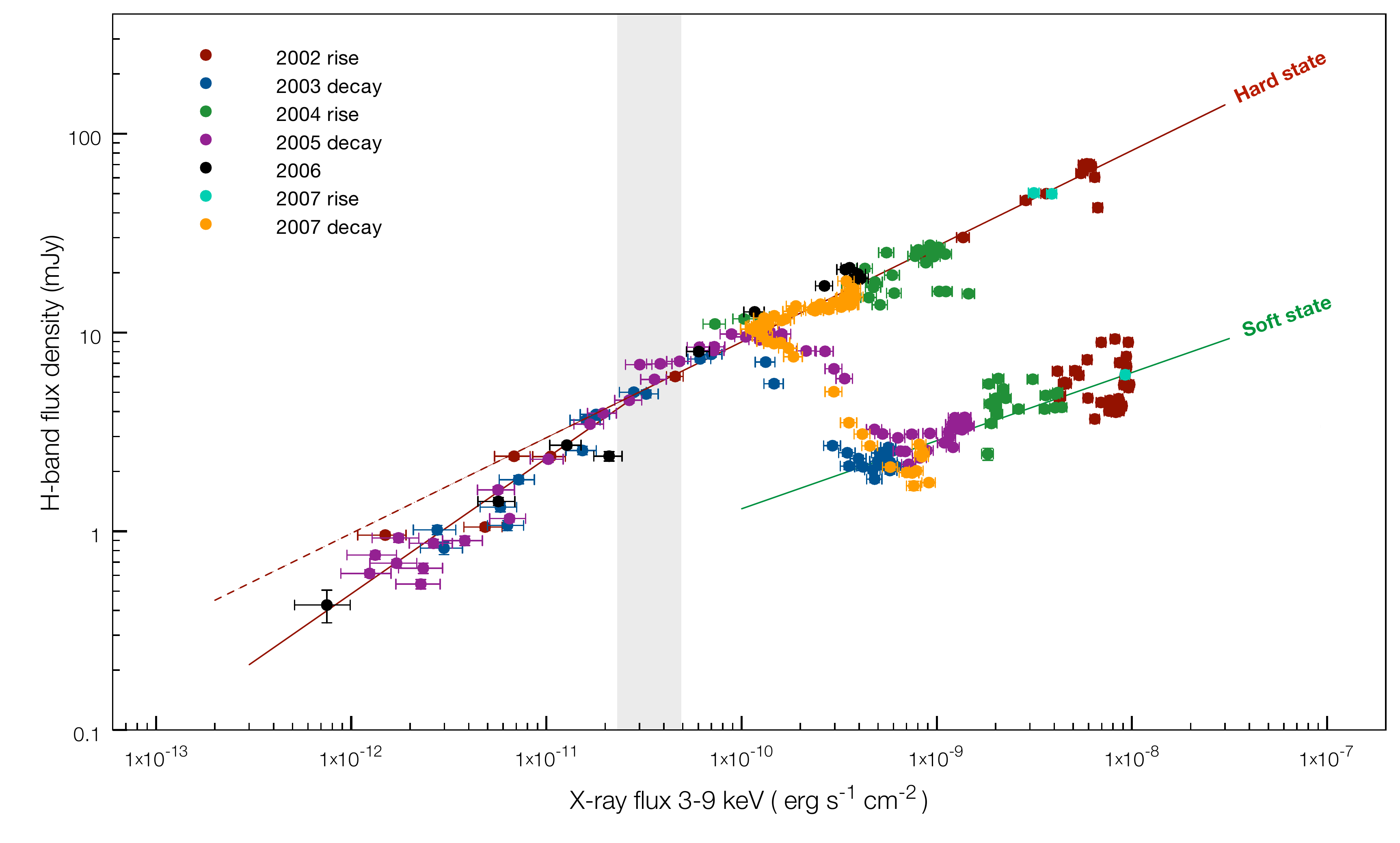} 
          \caption{Quasi-simultaneous infrared \textit{H-}band flux density versus 3-9 keV  X-ray flux during the four outbursts. Red and green continuous lines indicate the broken power-law fit to the hard state and the simple power law fit to the soft state data respectively. Grey zone indicates the 90 per cent confidence interval of the X-ray break flux $F_{\rm break}$.}
  \label{HX1}
  \end{figure*}

In order to study the connections between these three components,  we plot our quasi-simultaneous ($\Delta t \leqslant$ 1 day) infrared \textit{H} band 
(Figs. \ref{HVXhs}a, \ref{HX1}, \ref{HX2}) and optical \textit{V} band flux density (Fig. \ref{HVXhs}b) against the  3--9 keV X-ray flux for the four outbursts. 
In Fig. \ref{HVXhs}a, we show the evolution of the IR/X-ray correlation over the four outbursts with the data separated by X-ray state (hard state, intermediate state, soft state).  
In Fig. \ref{HX1}, we highlight  the difference between the rising and decaying phases of each outburst in order to search for existence or not of a `parallel track' effect (see below)
 in the IR/X-ray correlation. In Fig.  \ref{HX2}, the same data set  is separated by outbursts to compare their evolution with time.
 Finally, the optical/X-ray correlation separated by X-ray state is represented in Fig. \ref{HVXhs}b.

\subsubsection{Infrared vs. X-ray}

\begin{figure*}
 \centering
  \begin{minipage}{176mm}
     \includegraphics[width=88mm]{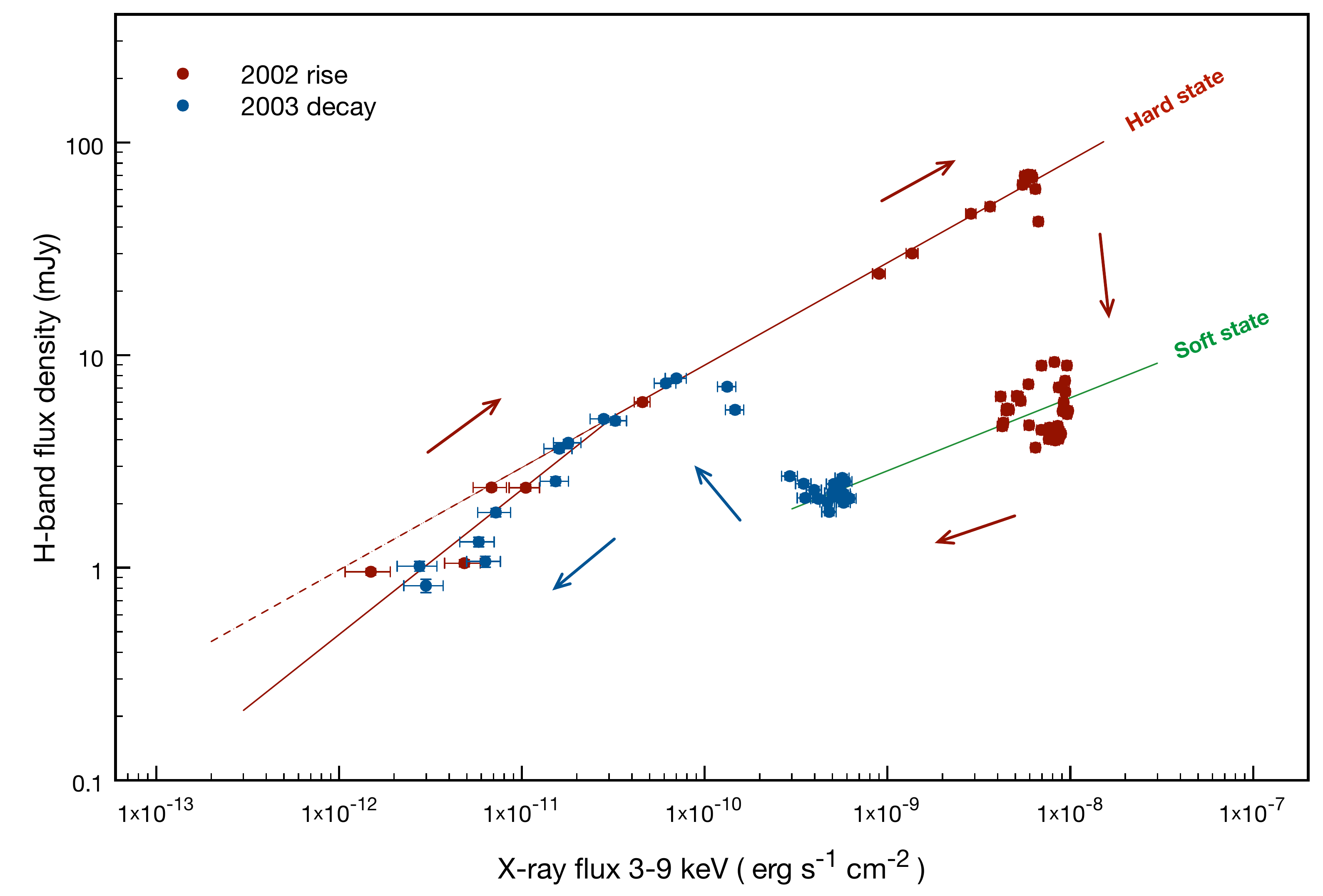}
     \includegraphics[width=88mm]{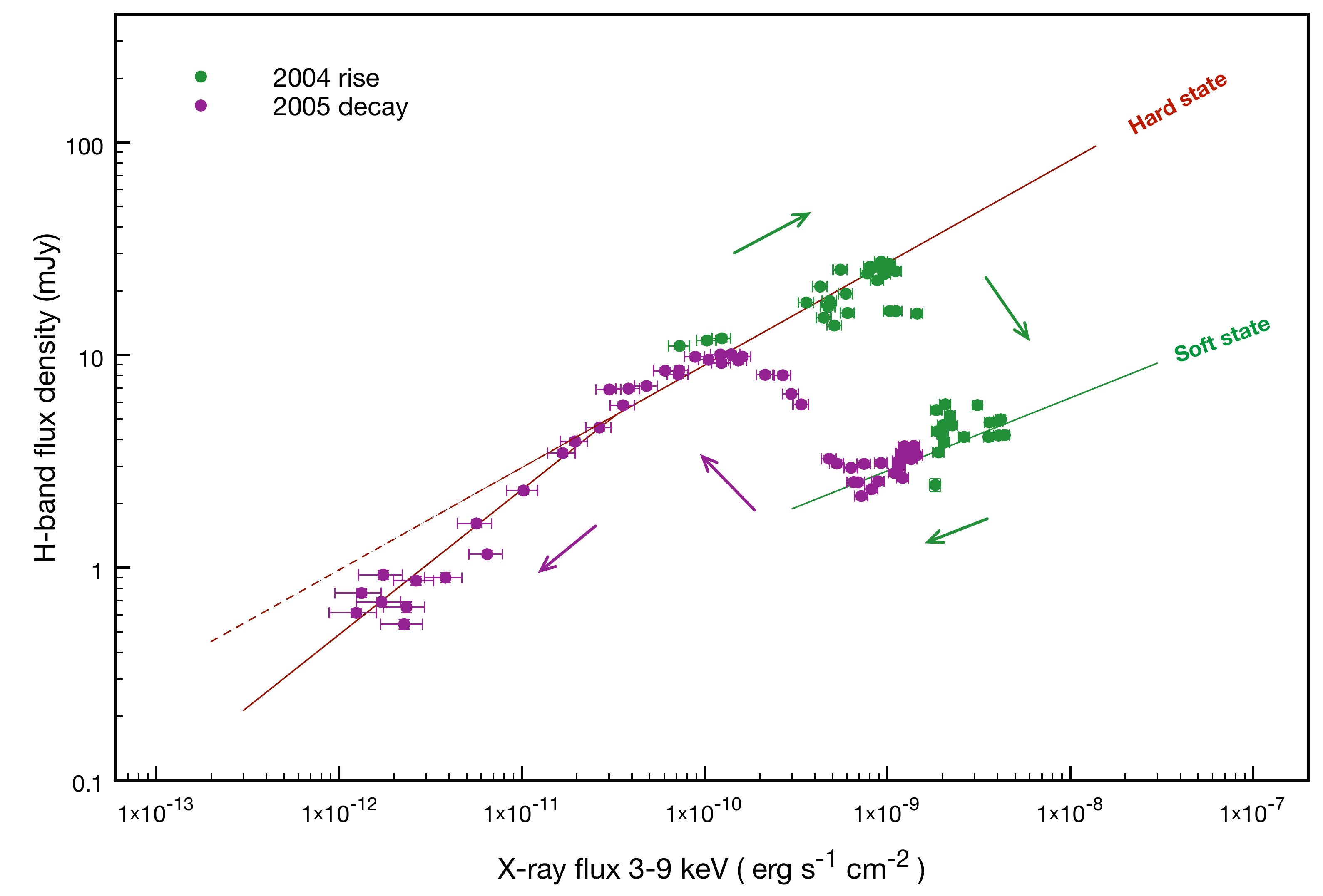} 
     \includegraphics[width=88mm]{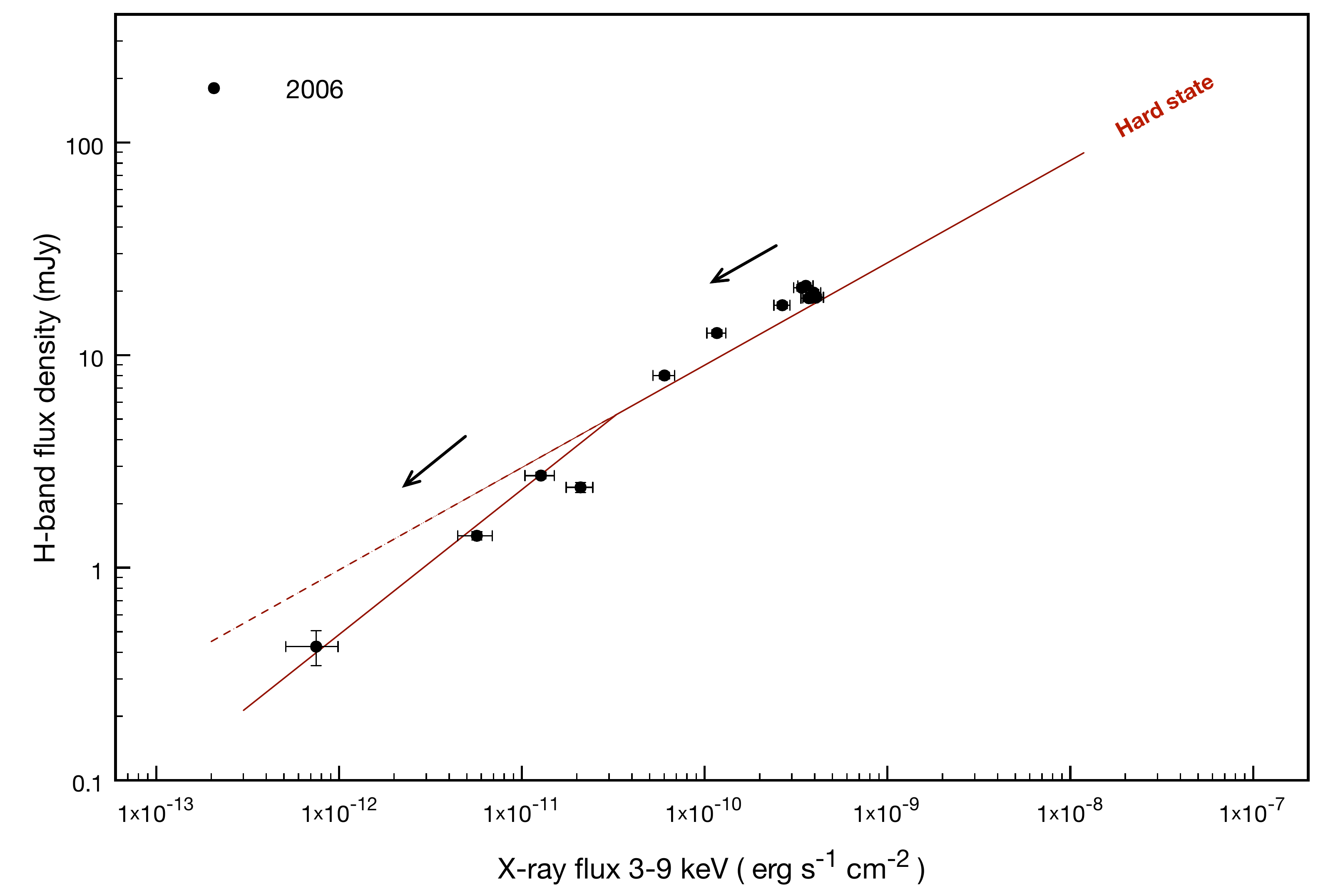} 
     \includegraphics[width=88mm]{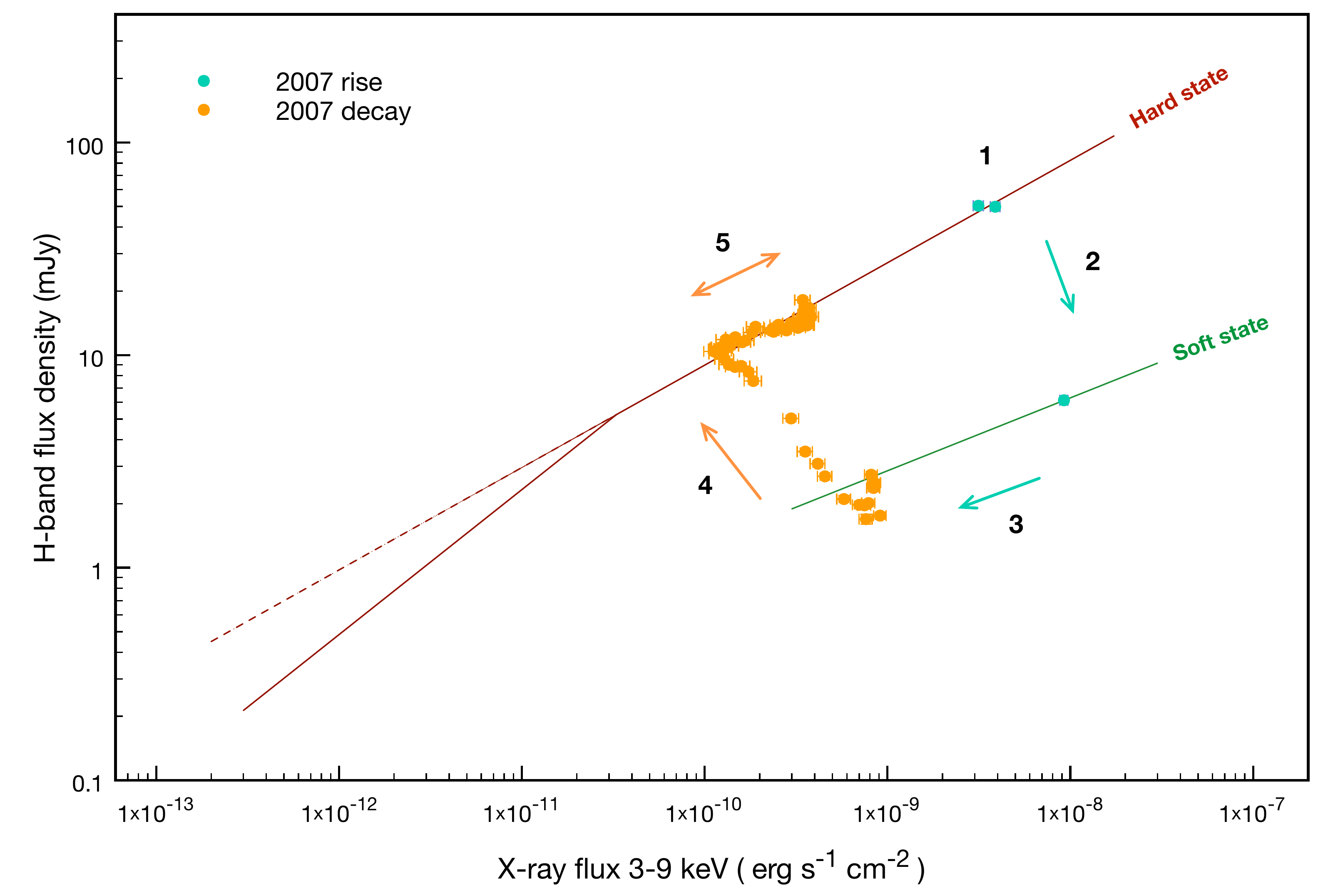} 
      \caption{Same as Fig. \ref{HX1} but with data set divided in outburst. Arrows are plotted to indicate the time evolution during the outburst.}
  \label{HX2}
  \end{minipage}
\end{figure*}

\begin{table*}
\begin{minipage}{162mm}
 \caption{Results of the non-linear power law fit ($F_{\rm{OIR}} = a \, F_{X}^{b}$) and broken power law fit ($F_{\rm OIR} = a \, F_{X}^{b_1}$ for $F_X \leqslant F_{\rm break}$ and $F_{\rm OIR} = a \, F_{X}^{b_2} \times F_{\rm break}^{b_2-b_1}$ for $F_X \geqslant F_{\rm break}$) of the OIR/X-ray correlations with $F_{\rm{OIR}}$ in mJy, $F_{X}$ and $F_{\rm break}$ in 
 $\rm{erg \,s^{-1} cm^{-2}}$. Errors are given at a 90 per cent level of confidence. }
 \label{fit}
 \begin{tabular}{llccccc}
  \hline
  Data set & X-ray state&$b$& $b_1$ & $b_2$& $a$&$F_{\rm break}$\\
  \hline
  \multirow{2}{3cm}{\textit{H-}band / 3--9 keV}
   &Hard state
   &--&$0.68 \pm 0.05$& $0.48 \pm 0.01$ &$\left(0.7^{+2.6}_{-0.5}\right) \times 10^8$&$\left(3.3^{+1.6}_{-1.0}\right) \times 10^{-11}$\\
   &Soft State&$0.34 \pm 0.01$ &--&--& $(3.5 \pm 0.6)  \times 10^3$&--\\
   \hline
   \multirow{2}{3cm}{\textit{V-}band / 3--9 keV}
   &Hard state& $0.44 \pm 0.01$&--&--& $(3.5 \pm 0.3) \times 10^5$&--\\
   &Soft State& $0.45 \pm 0.04$ &--&--&$\left(1.7^{+ 2.4}_{- 1.0}\right) \times 10^5 $&--\\
  \hline
 \end{tabular}
 \end{minipage}
\end{table*}

Fig.  \ref{HVXhs}a shows two distinct patterns for the hard state and the soft states that are connected by the intermediate-state data. 
In the hard state, IR and X-ray emissions are strongly correlated over four decades in X-ray flux. We note a possible deviation from a straight line: there seems to be a break in  the correlation to a steeper slope for data points at X-ray fluxes below $\sim 5 \times 10^{-11} \rm{\, erg \,s^{-1} cm^{-2}}$. To determine
the significance of this break, we fit the hard-state data in log space with a simple power law and a broken power law. 
The broken power law improves the quality of the fit with a $\chi^2$ of 584 (for 130 degrees of freedom (d.o.f.)) compared to a $\chi^2$ of 848 (for 132 d.o.f.)
 for the simple power law.  In addition, the confidence intervals at 90 per cent of the broken power law slopes $b_1$ and $b_2$ do not present any overlap ($0.64 \leqslant b_1 \leqslant 0.74$ and 
$0.47 \leqslant b_2 \leqslant 0.49$). We performed an F-test to compare the two model taking the simple power law model as the null hypothesis. We obtained a probability of $3\times10^{-11}$,  
which indicates that the deviation is indeed significant. We note however that both models give a reduced $\chi^2$ significantly greater than 1, so
there must be some source of intrinsic variability away from the power law relation that dominates over the observational errors.
 We found that the break occurs at $F_{\rm break} = (3.3^{+1.6}_{-1.0}) \times 10^{-11} \rm{\, erg \,s^{-1} cm^{-2}}$ in the 3--9 keV range. 
The corresponding bolometric (3--100 keV) flux has an average value of $1.1 \times 10^{-10} \rm{\, erg \,s^{-1} cm^{-2}}$. 
This gives a break luminosity $L_{\rm break} \sim 10^{-3} \, L_{\rm{Edd}}$ if we assume a black hole mass $M=5.8 \, M_{\sun}$ and a distance to the source $D=8$ kpc \citep{hynes03,hynes04}.

Likewise, in the soft state, a correlation exists between IR and X-ray emissions with a power-law slope of  $b = 0.34 \pm \,0.01$.
However, the correlation is weaker in the soft state (correlation coefficient: 0.84) compared to the hard state (correlation coefficient: 0.96 and 0.97 for the low- and high-luminosity portions, respectively). This supports the existence of  a change in the physical processes involved between these
 two states, as we expect the quenching of the compact jets and the thermal emission from the accretion disc to dominate the soft X-rays.

In Fig. \ref{HX2}, we note the similarities in the behaviour of  GX 339-4 over the four outbursts.
The source starts at low IR and X-ray fluxes along the lower flux power law correlation and then joins the upper hard-state track. The fluxes rise along this track until the hard-to-soft state transition is reached. From this point, the IR flux drops dramatically while the source reaches the soft state-track. It then evolves more randomly along this path, according to the flux variation during the soft state. IR and X-ray fluxes decrease until the soft-to-hard state transition where the IR flux strongly increases, leading back to the hard-state correlation. The fluxes then decline following the same path as the rising phase. We note however that only five points (from the 2002 outburst) at fluxes below $F_{\rm break}$ belong to a rising phase. This means that we cannot rule out the possibility that the source does not follow the steep hard-state track during the rising state of an outburst. It could be important to keep this in mind for further discussions.

\subsubsection{Optical vs. X-ray}\label{oirx}

The \textit{V-}band/X-ray data set (Fig. \ref{HVXhs}b) also exhibits two main correlation tracks for the hard state and the soft state. However, there are several differences with respect to the IR. If we fit the hard-state data with a simple or a broken power law as previously, the quality of the fit is not improved by the broken power-law model. This indicates that either no deviation is present in the \textit{V-}band/X-ray correlation in the hard state or it is too weak to be statistically significant. We note also that the drop in optical luminosity during the hard-to-soft state transition is weaker. These various differences may imply that in the optical bands, several other emission components, besides the jet emission, contribute or even dominate. As already suggested by previous studies \citep[e.g.][and references therein]{van-paradijs95,homan05,russell06}, these components may be blackbody emission of the accretion disc or reprocessed emission of the X-rays by the outer parts of the disc. 

We finally fit the hard- and soft-state data with the simple power-law model and we obtained a correlation slope of $b = 0.44 \pm \,0.04$ and $b = 0.45 \pm \,0.04$ for the hard state and the soft state respectively. All fit results for  IR/X-ray and optical/X-ray correlations are reported in Table \ref{fit}.

\subsection{Parallel tracks in the hard state } 

We focus here on the phenomenon pointed out by \citet{russell07}  for the IR/X-ray correlation of XTE J1550-564 during its 2000 outburst (phenomenon called `hysteresis' in this article) and for the radio/X-ray correlation of GX 339-4 over the 1997-2005 period by Corbel at al (in preparation). This is  the presence of quasi-parallel correlation tracks corresponding to the hard-state rising phase and the hard-state decaying phase of a same outburst\footnote{As discussed in Corbel et al. (in preparation), the two parallel tracks in the radio/X-ray correlation of GX 339-4 do not precisely correspond to a single track for the rise and a single track for the decay, unlike what is seen in XTE J1550-564 \citep{russell07}.}. According to Fig. \ref{HX1} and \ref{HX2}, there is no evidence for this particular phenomenon in the IR/X-ray correlations of GX 339-4. Indeed, the different hard state phases (rise and decay) of the four outbursts, share the same correlation track. Interestingly, we do observe two parallel correlation tracks in the radio/X-ray correlation during the 2002 and 2004 outburst of GX 339-4. This implies that, even if there is strong evidences for a jet origin of both the IR and radio emission in the hard state, they do not seem to be connected to the X-ray in the same way.

\subsection{Radio - IR connection}
\begin{figure}
    \includegraphics[width=84mm]{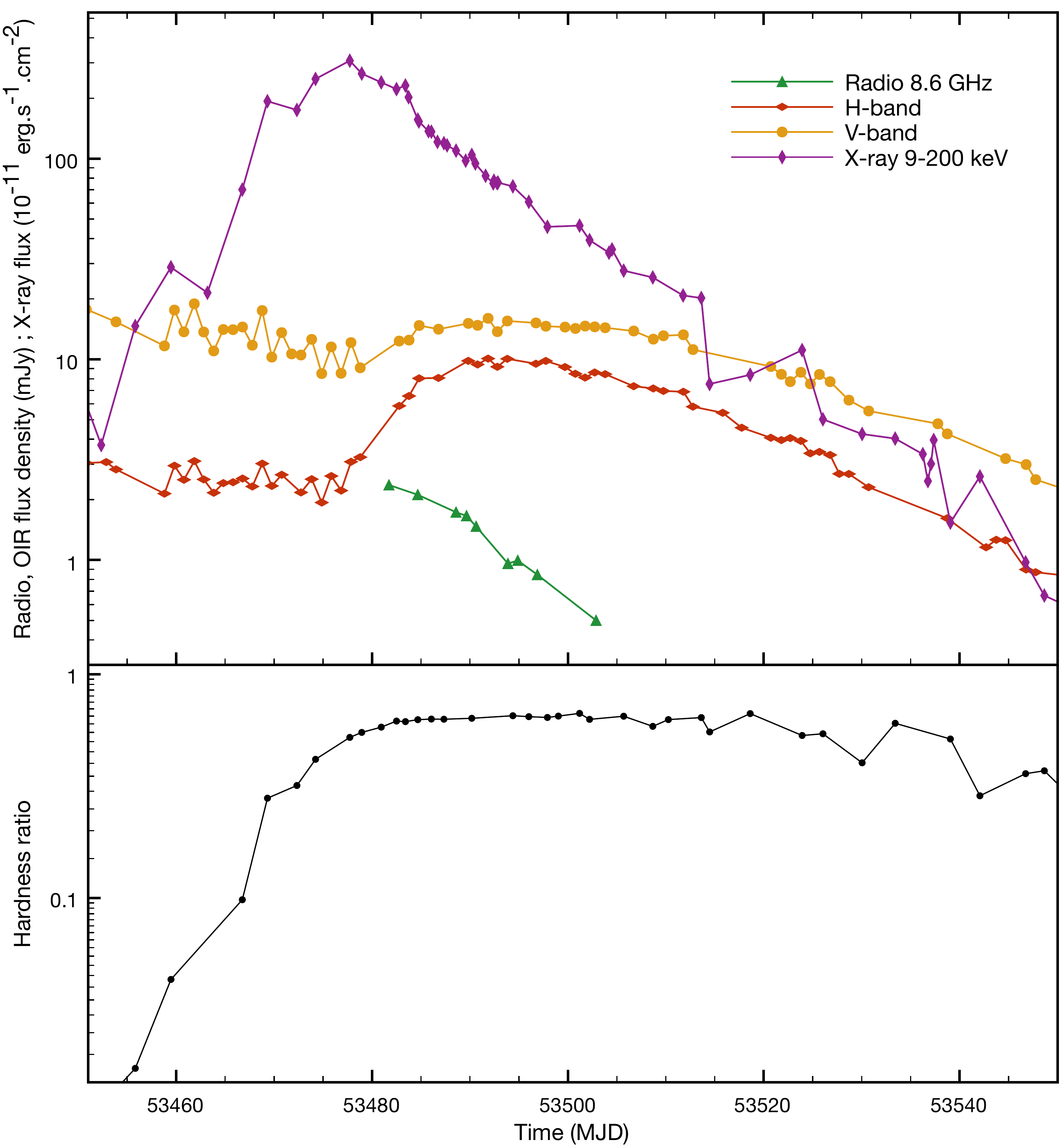} 
          \caption{\textit{Upper panel:} Radio 8.6 GHz (mJy), \textit{H-}band (mJy), \textit{V-}band (mJy) and X-rays 9--200 keV ($10^{-11} \  \rm{erg \; cm^{-2} \, s^{-1}}$) light curves at the beginning of the 2005 decaying hard state between MJD 53450 and MJD 53550. The radio flux densities have been divided by 2 to clarify the plot. \textit{Lower panel:} The corresponding X-ray hardness ratio as a function of time.}
  \label{lc}
  \end{figure}

\begin{figure*}
 \centering
  \begin{minipage}{180mm}
     \includegraphics[width=90mm]{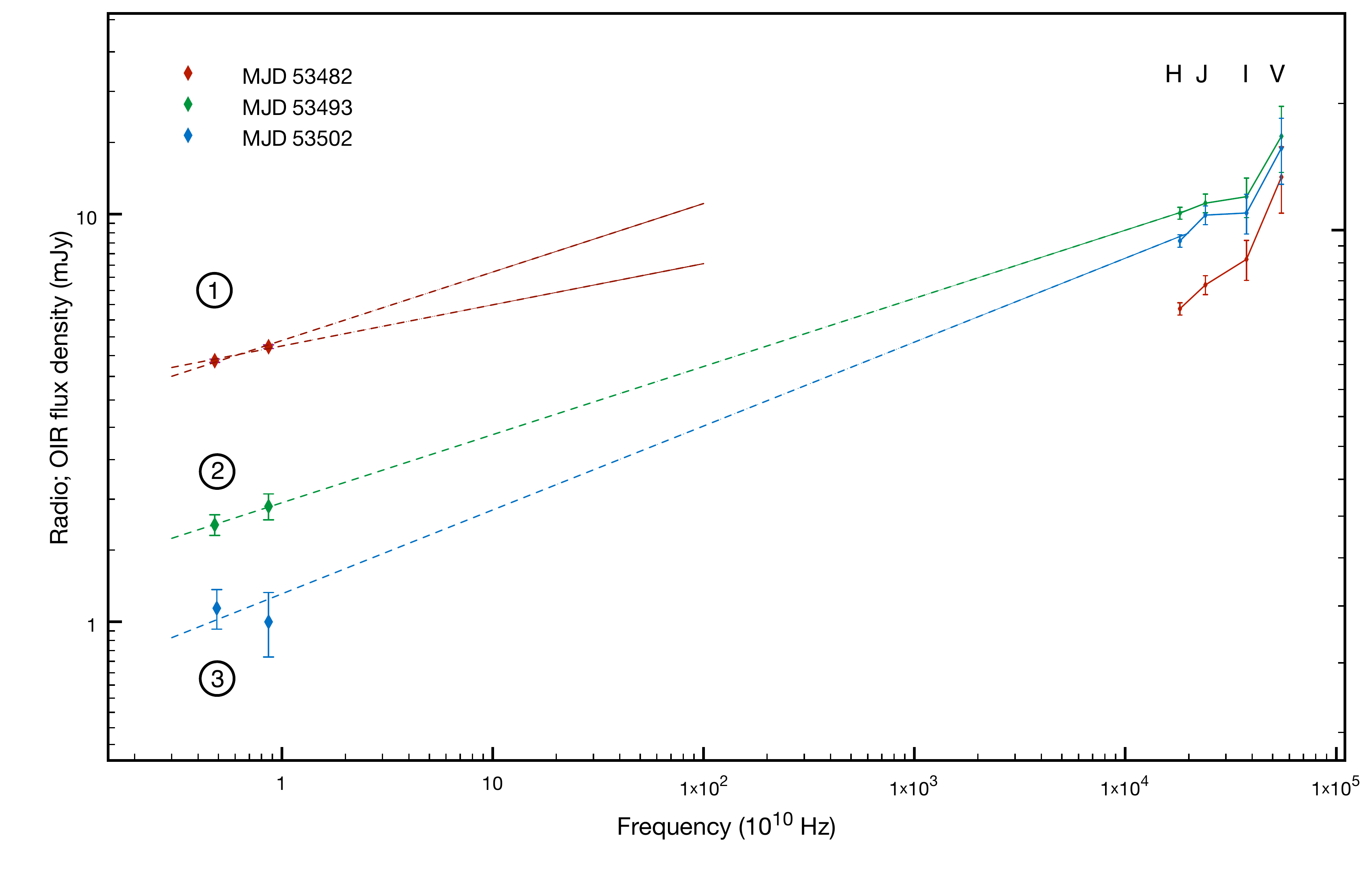} 
     \includegraphics[width=90mm]{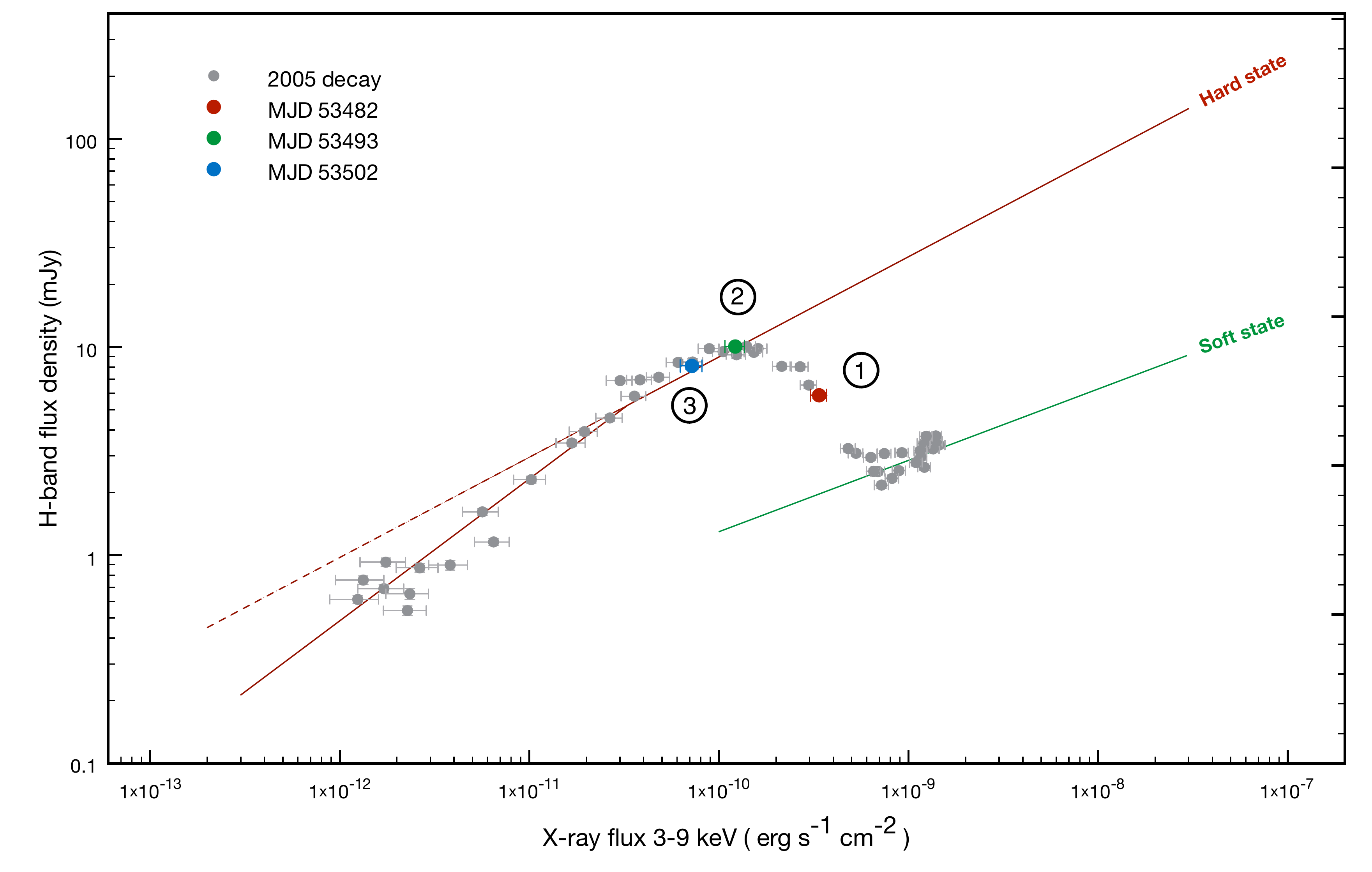} 
      \caption{\textit{(a) Left panel:} Radio to OIR SEDs during the onset of the compact jet in the 2005 decay. \textit{(b) Right panel:} The corresponding observations in the IR \textit{H-}band vs. X-ray 3--9 keV diagram. The continuous lines indicate the fits to the overall data set as already shown in Fig. \ref{HVXhs} and \ref{HX1}. }
  \label{sed}
  \end{minipage}
\end{figure*}

In Fig. \ref{lc}, we present the radio 8.6 GHz, the IR \textit{H}-band, the optical \textit{V}-band and the hard X-rays (9-200 keV) light curves during the beginning of the 2005 decaying hard state (for which we have the best radio coverage). A selection of  radio-to-OIR SEDs with their corresponding data points in the IR/X-ray diagram are then presented in Fig. \ref{sed}a and \ref{sed}b. 

At the beginning of the decaying hard state, we usually observe an increase of non-thermal hard X-ray emission \citep*{tomsick01}. In most cases, this is 
also accompanied by an increase in radio and IR emissions, suggesting  the onset of the compact jets once the source has settled down back in the hard state
  \citep{corbel00,jain01,buxton04,kalemci06}. Radio, IR and hard X-rays emissions continue to decrease while the source goes back to low luminosities. Fig. \ref{lc} is consistent with this description but we note an interesting delay between radio and NIR emissions.
Indeed, from the first radio observation (MJD 53481), radio and hard X-ray fluxes are already decreasing while the IR peak flux occurs about 12 days later ($\sim$ MJD 53493). The evolution in optical is similar to the IR, with a flux increase of $\sim$7 mJy in \textit{V}-band and $\sim$8 mJy in \textit{H}-band.

Fig. \ref{sed}a represents three selected radio-to-OIR SEDs corresponding to the dates of the second radio observation (no OIR data were available during the first one), the peak of the IR flux and the last radio observation. The first SED (MJD 53482) displays a radio spectrum with a power-law slope $\alpha = 0.13 \pm 0.02$ consistent with the typical optically thick synchrotron emission of a self-absorbed compact jet (for a flux density S$_{\nu} \propto \nu^\alpha $). 
We note that the OIR lies significantly below the extrapolation of the radio spectrum. As the radio emission originates from the optically thick portion of the compact jet, the 
transition frequency to the optically thin regime should be located at lower values than the \textit{H}-band frequencies. However, optically thin synchrotron emission cannot be dominant in OIR since the OIR spectrum increases with frequency. This rather suggests a contribution from the outer parts of the accretion disc (disc blackbody, X-ray reprocessing). In the IR/X-ray diagram (red point in Fig. \ref{sed}b), the corresponding data point lies in the transition zone between the soft-state and the hard-state correlation tracks.
The second SED, 11 days later (MJD 53493), shows a decrease of the radio flux by a factor $\sim3$, and an increase in IR and optical by a factor $\sim2$ and $\sim0.5$ respectively. We can see that the \textit{H-}, \textit{J-} and \textit{I-} bands fluxes are now consistent with an extrapolation of the radio spectrum by a power law with a slope $\alpha = 0.17 \pm 0.01$. The connection of the radio and IR spectra would indicate an increase of the turnover frequency to values greater than the \textit{H-}band frequencies. The shape of the OIR spectrum still suggests a significant thermal contribution from the disc emission. On Fig. \ref{sed}b (green point), the corresponding IR/X-ray data point has reached the main hard-state correlation. The last SED (MJD 53502) displays features similar to the previous one. The IR spectrum is  consistent with an extrapolation of the radio spectrum with 
an inverted power law of spectral index   $\alpha = 0.21 \pm 0.02$  and the optical (\textit{V},\textit{I}) spectrum still suggests the contribution of the disc emission. These three SEDs seem to indicate that the radio to infrared spectra get more inverted as the flux (radio and X-ray) decreases.

\section{Discussion}

\subsection{OIR/X-ray correlations in the hard state.}

\subsubsection{Are the IR/X-ray correlations consistent with synchrotron self-Compton X-rays ?}

During the hard states of each of the four outbursts of GX 339-4 between 2002 and 2007, we found a strong correlation of the form $L_{\rm{IR}} = a \, L_{X}^{b}$ between the infrared \textit{H-}band luminosity, $L_{\rm IR}$, and the 3--9 keV  X-ray luminosity, $L_X$, with a slope $b_2 = 0.48 \pm 0.01$, which breaks into $b_1 = 0.68 \pm 0.05$ at X-ray fluxes below $F_{\rm break} =  \left(3.3^{+1.6}_{-1.0}\right) \times 10^{-11} \rm{\, erg \,s^{-1} cm^{-2}}$.

Previous OIR studies of BHXBs in outbursts \citep{jain01,corbel01,corbel02,buxton04,homan05,russell06,russell07} have presented evidences for a compact jet origin of the NIR emission in the hard state. We also know by spectral analysis that the  3--9 keV X-ray emission is dominated in the hard state by a non-thermal component whose origin is usually attributed to inverse Compton emission from a hot plasma, a corona surrounding the compact object or the base of the compact jet. Our results confirm the tight connection between these two frequency domains. But thanks to the extensive data set collected during a period of 5 years and over four outbursts of a given source, it allows us  to constrain the correlation index and 
 to reveal some additional behaviour that has not been seen in a smaller data set or in a sample combining several sources, and hence, introducing more scattering due to uncertainties in distance estimates. On the other hand, the results and interpretations presented here  are only valid for one source, and 
 would need to be validated for other sources before extrapolating its universality. 

Regarding the correlation indices in the hard state, we know that several models have explained the values found in previous IR/X-ray or radio/X-ray studies
 (e.g. \citealt{merloni03,markoff03,markoff05}; \citealt*{yuan05}). So the interpretation presented below is not unique but is an attempt to explain in a consistent way the results outlined in the previous section and in particular the indices of the power-law fit and the break around $ \sim 10^{-3} \, L_{\rm{Edd}}$.

\begin{figure}
    \includegraphics[width=84mm]{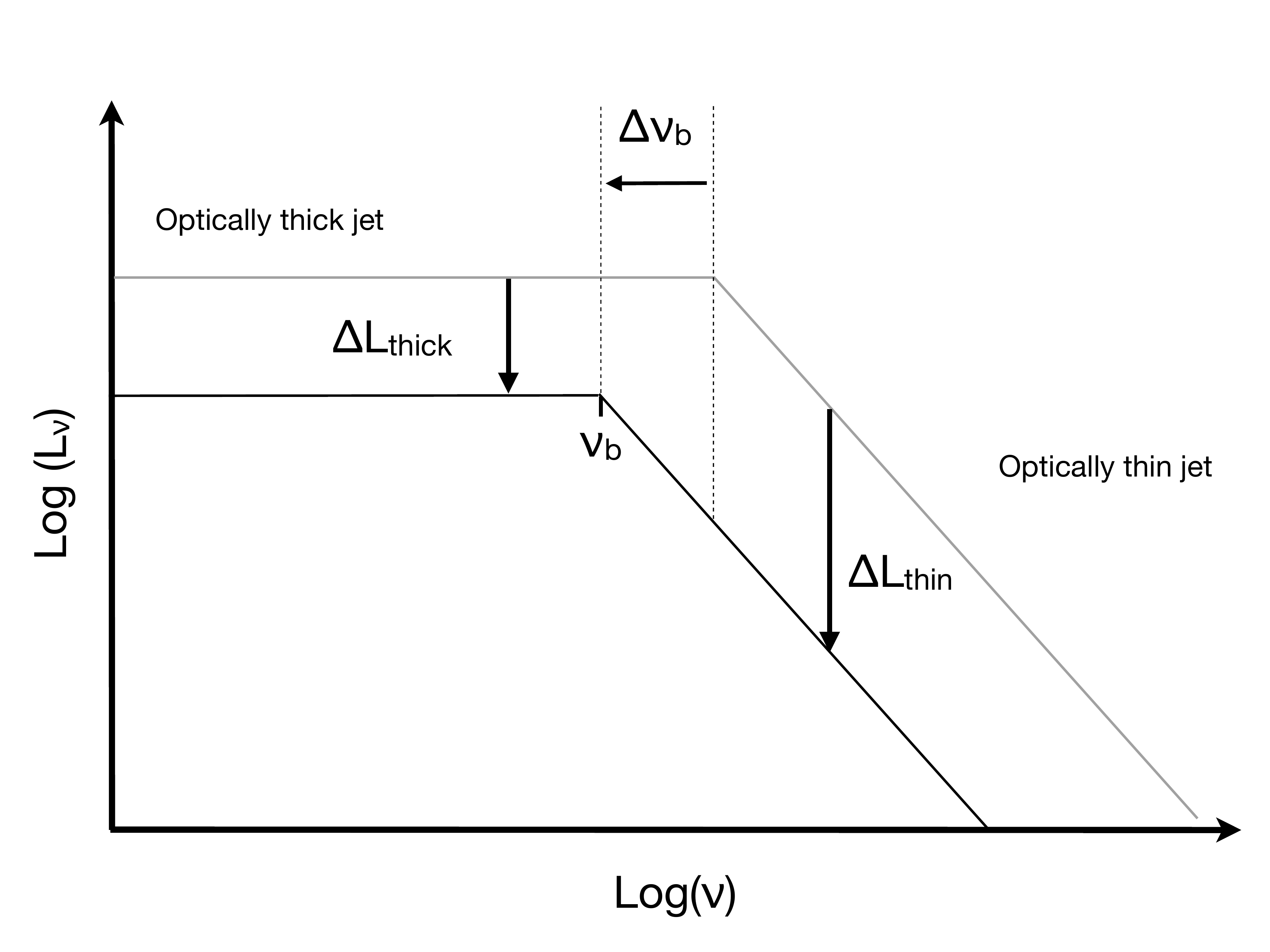} 
          \caption{A schematic jet spectrum which illustrates that for a given variation of the jet power, the corresponding variation of the luminosity will be stronger in the optically thin regime compared to the optically thick portion, due to the variation of the break frequency.}
            \label{break}
  \end{figure}

As already suggested by \cite{markoff05}, the non-thermal hard X-ray emission could arise from inverse Comptonisation of the synchrotron photons produced in the compact jets, by the plasma producing the synchrotron emission. This synchrotron self-Compton (SSC) process, implies a dependence of the X-ray luminosity,  $L_X$, with the jet power,  $ Q_{jet}$ (\citealt{falcke96a}, see also discussion in \citealt*{corbel08} for V404 Cyg) as: \begin{equation}
 L_X \propto Q_{jet}^{11/4}
 \label{eq2} 
\end{equation}
From the standard conical jet model, we expect the monochromatic luminosity at a given frequency $L_{\nu}$ to scale with the frequency as $L_{\nu} \propto \nu^{\alpha}$. If the frequency lies in the optically thick (flat or inverted) part of the spectrum, we usually observe $\alpha=$ 0--0.15 and if it lies in the optically thin part, we expect $\alpha= -0.5$ to $-0.7$. If we call $\nu_b$ the break frequency between the optically thick and the optically thin portion of the spectrum, we get:
\begin{equation}
L_{\nu} = L_{\nu_b} \left(\frac{\nu}{\nu_b}\right)^{\alpha}
\label{eq3}
\end{equation}
As described in \cite{falcke96a} and \cite{markoff03}, it follows from simple analytic arguments that $L_{\nu_b}  \propto Q_{jet}^{17/12}$ and $\nu_b \propto Q_{jet}^{2/3}$ and hence we get the monochromatic luminosity at a given frequency:
\begin{equation}
L_{\nu} \propto Q_{jet}^{\frac{17}{12}-\frac{2}{3} \alpha}
\end{equation}
Using equation (\ref{eq2}), we get:
\begin{equation}
L_{H} \propto L_X^{\left( \frac{17}{12}-\frac{2}{3} \alpha \right) \times \frac{4}{11} }
\end{equation}
where $L_H$ is the \textit{H-}band luminosity.
If we consider the typical ranges of values of $\alpha$ given above, we obtain $b =$ 0.48--0.52 if the \textit{H-}band lies on the optically thick part ($\alpha=$ 0--0.15) and  $b =$ 0.63--0.68 if it lies on the optically thin part
($\alpha= -0.5$ to $-0.7$). These values are fully consistent with the indices we derived for the hard state: $b = 0.48 \pm 0.01$ for the high-flux part and $b = 0.68 \pm 0.05$ for the low-flux part. As the break frequency varies with jet power ($\nu_b \propto Q_{jet}^{2/3}$), this suggests that the \textit{H-}band is located on the optically thick part of the spectrum for $L_X \ga  10^{-3} \, L_{\rm{Edd}}$ and on the optically thin part when $L_X$ goes below $\sim 10^{-3} \, L_{\rm{Edd}}$. The results are therefore consistent with an SSC origin of the X-rays and it is interesting that this interpretation could explain in a consistent way the values of the observed indices and the presence of the break in the correlation (but see section \ref{adaf} for an alternative interpretation). Fig. \ref{break} illustrates the fact that, for a given variation of the X-ray luminosity, the corresponding variation in the IR will be more important if it lies on the optically thin part of the jet spectrum than on the optically thick part, and hence, will give a much steeper slope in the correlation.

\subsubsection{Break frequency}

Under the previous assumptions on the origin of the break in the correlation, we can calculate an expression giving a rough estimate of the break frequency as a function of the monochromatic luminosity $L_{\nu}$. As presented above, in a standard jet model the luminosity depends on the jet power as  $L_{\nu_b}  \propto Q_{jet}^{17/12}$ and the break frequency as $\nu_b \propto Q_{jet}^{2/3}$. Hence, we find $L_{\nu_b}  \propto \nu_b^{17/8}$, and we can write:
\begin{equation}
L_{\nu_b} = a\, \nu_b^{17/8}, 
\label{eq6}
\end{equation}
where $a$ is the normalisation factor.  The broken power-law fit gives a value of the \textit{H-}band flux where the break in the correlation occurs of $F_{H} \simeq 5.2 \pm 1.5$ mJy. At this flux density, we therefore have $\nu_b \approx \nu_H$, with $\nu_H = 1.18 \times 10^{14} \; {\rm Hz}$ the effective frequency of the \textit{H-}band filter. Consequently, by converting the \textit{H-}band flux density $F_{H}$ into the monochromatic luminosity $L_{H}$ at the frequency $\nu_H$, we obtain the following normalisation factor for an assumed distance to the source of 8 kpc:
\begin{equation}
a \simeq \left(5 \pm 1.5\right) \times 10^{-10} \; \left( \frac{D}{8\, kpc}\right)^{2} \; {\rm erg \; s^{-1} \, Hz^{-\frac{25}{8}} \,}
\label{eqa}
\end{equation}
Using equation (\ref{eq3}) and (\ref{eq6}) we can finally express the break frequency as a function of a measured monochromatic luminosity $L_{\nu}$ in the jet spectrum, its corresponding frequency $\nu$ and the spectral index $\alpha$ where this frequency is located:
\begin{equation}
\nu_b = \left(\frac{L_{\nu}}{a \, \nu^{\alpha}}\right)^{\frac{1}{ \frac{17}{8} -  \alpha}},
\end{equation}
with the normalization constants stated above. Note that the above expression has been derived without any assumption on the physical origin of the X-ray emission in the hard state. 
Additionally, we would like to point out the work of \citet{nowak05} on the scaling of the break frequency with the soft X-ray flux in GX 339-4. The authors infer the location of the break frequency by fitting the
 radio and X-ray data by a doubly broken power law which implicitly assumes that the soft X-ray emission (3--9 keV) is dominated by direct synchrotron emission from the jet. However, 
 since we do not adopt the same assumptions, we cannot compare consistently their work with our results.

\subsubsection{Is the IR/X-ray correlation consistent with an accretion flow origin of the X-rays ?}\label{adaf}

 To explain the fundamental plane of black hole activity, \citet{merloni03} tested several classes of accretion flow model that could be responsible for the X-ray power law emission. Based on the initial work by \citet{heinz03},
 they provided analytical expressions to determine the correlation coefficient between radio and X-ray flux that we can use to test these models with our data. We have selected the two classes of models the authors identified as consistent with their results: the ADAF case and a more general class of  radiatively inefficient, mechanically cooled accretions flows that are modified by convection or outflows and where we expect the X-ray spectrum to be dominated by bremsstrahlung emission. 

Note that in the following calculations, we used a power law index of the electron distribution $p=$ 2--2.4. For the ADAF case with an optically thick synchrotron IR spectral index\footnote{Note that the spectral index $\alpha$ used by \citet{merloni03}, is defined with the following convention: $F_{\nu} \propto \nu^{-\alpha}$, which is the opposite of the one adopted in this paper. } ($\alpha=$ 0--0.15), we obtain $b =$ 0.56--0.62. In the optically thin case ($\alpha= -0.5$ to $-0.7$) we get $b =$0.76--0.80. The bremsstrahlung dominated case gives slightly higher values of $b:$ for  $\alpha=$ 0--0.15 we obtain $b =$ 0.66--0.71 and for $\alpha= -0.5$ to $-0.7$ we get $b =$ 0.88--0.92. 

Consequently, in the frame of an accretion flow origin of the X-rays, an alternative interpretation of our results would be to consider that the \textit{H-}band is constantly located in the optically thick part of the jet spectrum ($\alpha=$ 0--0.15) and that the X-ray emission originates from an ADAF at high luminosities that becomes bremsstrahlung dominated at low luminosities. This would give $b =$ 0.56--0.62 at high luminosities, which is roughly consistent with our derived index $b = 0.48 \pm 0.01$, and would become $b =$ 0.66--0.71 at low luminosities, fully consistent with our $b = 0.68 \pm 0.05$.

\subsubsection{Optical/X-ray correlation in the hard state}

As reported in Section \ref{oirx}, in the hard state the optical emission shows a correlation with the X-rays in the form of a power law with slope $b=0.44 \pm 0.01$. Unlike the IR versus X-ray correlation, it does not show any significant break in the correlation. Following equations (\ref{eq6}) and (\ref{eqa}) and assuming that the synchrotron emission from the jet dominates the optical domain, we would expect the break to occur at a \textit{V-}band flux around 130 mJy, which is above the maximum flux reached in our data set (Fig. \ref{HVXhs}b). Consequently, if the optical emission is related to the compact jets optically thin synchrotron, we would indeed expect a simple power law to fit the entire optical/X-ray hard-state data with a correlation index $b$ close to 0.68 as seen in IR below the break. However the shape of the OIR SEDs and the weaker gap in \textit{V-}band luminosity compared to the IR luminosity between the hard and the soft state, rather suggest that another emitting component than the compact jet dominates the optical domain as already suggested by  e.g. \citet{van-paradijs95,homan05,russell06}. This component could be the accretion disc or the companion star and in both cases it can be direct blackbody emission or reprocessing of the X-rays. \citet{homan05} already showed that we can basically rule out the secondary star as the dominant source of OIR emission in GX 339-4 during outburst. The accretion disc is therefore the most likely dominant source in optical in both hard and soft state. Consequently, the observed correlation between optical and X-ray (3--9 keV) point out the connection between the outer parts of the accretion disc and the non thermal emission in X-ray which dominates the 3--9 keV range in the hard state. Regarding the origin, irradiated or not, of the outer disc emission, we can see on Fig. \ref{lc} that the delay between the hard X-rays and OIR peaks is around 12 days. If the hard X-rays are indeed the source of irradiation and the OIR the reprocessed emission, a 12 days delay is highly inconsistent with a reprocessing mechanism.

Under the above assumptions, the loss of optical emission during the hard-to-soft state transition gives an estimate of the synchrotron contribution in optical of $40$ per cent in average.

\subsection{OIR/X-ray correlations in the soft state}
In the soft state, the drop in IR flux is indicative of the quenching of the jet emission. The OIR SEDs, that can be well  fitted with  simple power laws suggest that the remaining IR emission in the soft state originates from the outer parts of the accretion disc. In that case, the emission may be dominated by thermal emission from a viscously heated disc or an X-ray heated disc. 

Let us consider a simple model of a steady thin disc with the temperature scaling with the radius as $T\propto r^{-n}$, where $n = 3/4$ for a standard disc and $n=1/2$ for an irradiation dominated disc \citep*{frank02}. Depending on the emission regime of the disc spectrum, we have the following relation between the monochromatic luminosity at a given frequency 
$\nu$ and the temperature: 
\begin{enumerate}
\item In the Rayleigh-Jeans (RJ) tail: 
\[
 \quad L_{\nu_{\rm RJ}} \propto T \nu^2
 \]
\item in the `flat' part\footnote{Portion of the disc spectrum between the Rayleigh-Jeans tail and the Wien cutoff, usually associated with a spectral index $\alpha=1/3$ in the case of a viscously heated disc.} of the disc spectrum:
\[
\quad L_{\nu_{\rm flat}} \propto T^{\frac{2}{n}}\; \nu^{\,3-\frac{2}{n}}
\]
\end{enumerate}
If we assume, that $L_X \propto T_{in}^4 \propto T^4$, where $L_X$ is the X-ray luminosity of the disc and $T_{in}$ its inner temperature, we obtain:
\begin{equation}
L_{\nu_{\rm RJ}}\propto L_X^{1/4} \quad {\rm and} \quad L_{\nu_{\rm flat}} \propto L_X^{1/2n}
\end{equation}
Therefore, we expect $L_{\nu_{\rm flat}} \propto L_X^{2/3}$ for a standard disc and $L_{\nu_{\rm flat}} \propto L_X$ for an irradiated disc. The RJ tail relation remains the same in both cases, $L_{\nu_{\rm RJ}}\propto L_X^{1/4}$. Table \ref{disc} summarises the expected power law indices depending on the emission regime and the disc model. Our derived index $b = 0.34 \pm 0.01$ for the IR/X-ray correlation in the soft state is thus inconsistent with the \textit{H-}band lying on the `flat' part of a standard or irradiated disc and is only marginally consistent with the expected relation for the RJ tail. 

\begin{table}
\centering
 \caption{Expected power law indices, $b$, in the relation $L_{\nu} \propto L_X^b$, depending on the emission regime and the disc model.}
 \label{disc}
 \begin{tabular}{ccc}
  \hline
  Emission regime & Viscous disc & X-ray heated disc\\
  \hline
  Rayleigh-Jeans tail & $b = 0.25$ & $b = 0.25$ \\
  Flat part & $b = 0.66$ & $b = 1$\\ 
  \hline
 \end{tabular}
\end{table}

However, this index could be explained by considering a frequency lying at the transition between the RJ part and the flat part of the disc spectrum. Indeed, the correlation index $b$, expected from the analytical relations given above, is inversely proportional to the spectral index $\alpha$ of the disc spectrum. Given that this spectral index decays from the RJ tail to the `flat' part, we would expect the corresponding correlation index to increase from $b=0.25$ to $b=0.66$ or $b=1$ for a standard or irradiated disc respectively. Consequently, our derived index $b = 0.34 \pm 0.01$ could be interpreted as the \textit{H-}band lying in the transition zone from the RJ tail to the `flat' part of the disc spectrum. 

\begin{figure}
    \includegraphics[width=84mm]{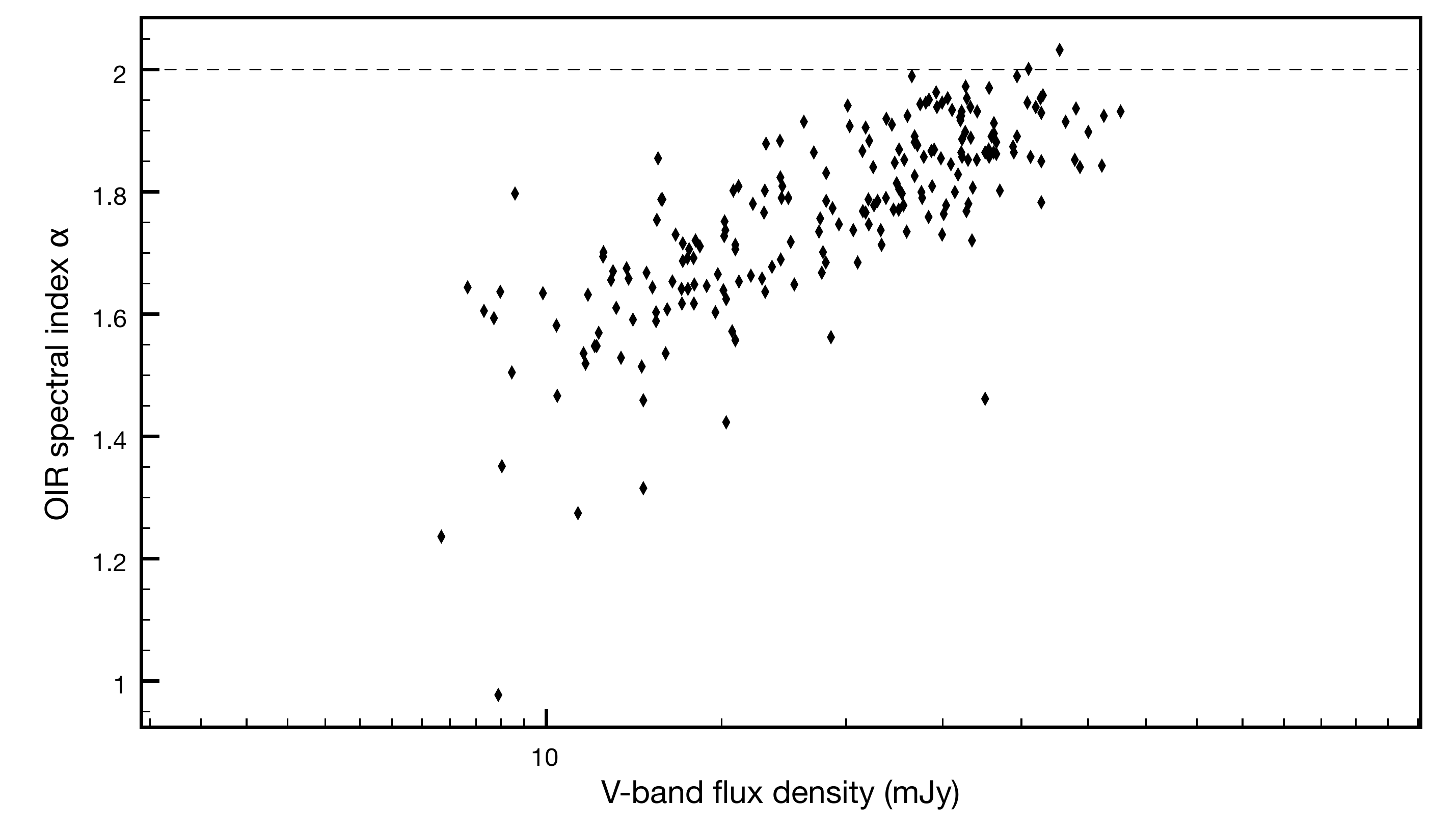} 
          \caption{OIR spectral index against \textit{V-}band flux (mJy). The indices are derived from the \textit{H-} to \textit{V-}band SEDs during the soft state phases only. Dashed line indicate the Rayleigh-Jeans tail value of the spectral index. Error bars on the indices and the \textit{V-}band flux are not plotted for clarity purpose and correspond to an average value of $\pm 0.2$ for the spectral index and $\pm 20$ per cent for the \textit{V-}band flux. Note however, that these errors are dominated by the uncertainty on the optical extinction and will therefore affect all the data in the same way without changing the global trend.}
  \label{alpha}
  \end{figure}

This interpretation also implies that the correlation index between the \textit{V-}band luminosity and the X-ray luminosity in the soft state has to be greater than 0.34 since the \textit{V-}band should be further in the transition zone than the \textit{H-}band. This is, indeed, what we observe: $b=0.45\pm 0.04$ for the optical/X-ray correlation in the soft state. 
The idea that the \textit{H-} to \textit{V-}bands are located in the transition from the RJ  to the `flat' part,  is also supported by the spectral indices derived from the OIR SEDs. Fig. \ref{alpha} shows the evolution of the \textit{H-}band to \textit{V-}band spectral index with the optical flux. We can see that the spectral index increases with the optical flux from $\sim 1$ to $\sim 2$. Since we expect the frequency of the break between the RJ and the `flat' part to increase with mass accretion rate, and thus, with optical flux, this suggest that the OIR bands are `moving' from the transition zone to the RJ tail with the spectral index reaching, at the highest fluxes, the expected value $\alpha = 2$ for the RJ tail.

However, in the theoretical frame described above, our derived indices do not allow us to discriminate between a viscously heated or an X-ray heated disc dominating the OIR emission in the soft state. 
This conclusion slightly differs from those obtained by \citet{russell06,russell08b} where evidences are provided for an irradiated disc dominating the OIR emission in the soft state. These differences arise from the way we derive the expected correlation indices between OIR and X-ray luminosity in the case of an irradiated disc. Based on \citet{van-paradijs94}, \citet{russell06,russell08b} expect $L_{\rm OPT} \propto L_X^{0.5}$ for an irradiated disc, whereas we expect $L_{\rm OPT} \propto L_X^{0.25 - 1}$ using the relations given in e.g. \citet{frank02}. This implies that, unlike \citet{russell06}, we cannot conclude on the reprocessed nature of the optical emission in the soft state. We think that further investigations, especially with broad band SED fitting and time lag studies, are necessary to clearly address this issue  \citep*[see the discussion in e.g.][and references therein]{gierlinski09}. We note, however, that apart from this irradiated disc interpretation, our specific results on GX 339-4 are consistent with the global interpretations of \citet{russell06,russell08b}.

\subsection{Radio - IR connection}
Fig. \ref{lc} and \ref{sed} shows the evolution of the OIR, radio and hard X-ray emissions during the formation of the compact jets. We first note the $\sim 12$ days lag of the OIR peak against the hard X-rays. This lag has been already observed in several BHXBs during outburst decay \citep{kalemci05,kalemci06} even if it is usually of the order of $\sim$ 3 days. These has been interpreted as the time needed for an optically thin environment (corona, jet base) to be formed before launching a jet. We note also that the OIR peak follows the radio by at least 10 days. If the increase of the OIR emission is indeed due to the formation of the compact jets, this delay implies that the onset of the jets synchrotron emission starts at low frequency. This is natural since we must have a phase transition where the energy density (magnetic, kinetic) of the jet has to increase before it reaches its `stable' structure and this would first produce radio emission before OIR as seen in Fig. \ref{lc}. This idea is also supported by the SEDs shown in Fig. \ref{sed}a which suggest an increase of the break frequency of the jet spectrum between MJD 53482 and MJD 53493.
However, an important question to address is whether the observed delay between radio and OIR is compatible with the time scales involved in a jet formation. But this requires a detailed model for the creation of a jet which is beyond the scope of this paper.

\section{Conclusions}

Using an extensive broadband data set, we compared the connections between X-ray, OIR and radio properties in the BHXB GX 339-4. We can summarise our main conclusions as follows:

\begin{enumerate}
\item We first see a strong power law correlation between IR and X-ray fluxes in the hard state with the presence of a break in the correlation slope below an X-ray luminosity around $\sim 10^{-3} L_{\rm Edd}$.
This break and the values of the correlation indices can be interpreted in a consistent way if we consider an SSC origin of the X-rays in the hard state and the variation of the break frequency of the jet spectrum. 
However, we note that our results are not inconsistent with an accretion flow origin of the X-ray if we consider that the X-ray emission originates from an ADAF at high luminosities that becomes bremsstrahlung dominated at low luminosities.
\item  Optical and X-ray emissions in the hard state display a correlation as well but suggest that the outer parts of the accretion disc dominate the optical emission in the hard state.
\item In the IR/X-ray correlation in the hard state, we do not detect any parallel tracks similar to what is seen for XTE J1550-564 \citep{russell06} or in the radio/X-ray correlation of GX 339-4 (Corbel et al. in preparation).
\item In the soft state, the correlations between the OIR and the X-rays indicates also a disc origin of the OIR emission and suggest that the \textit{H-}band and \textit{V-}band are located in the transition zone between the Rayleigh-Jeans tail and the `flat' part of the disc spectrum.
\item We compared hard X-ray, optical, infrared and radio light curves during a selected transition to the hard state and found that the changes in the OIR emission follow the changes in hard X-ray and radio. These delays  rule out a reprocessing mechanism as the origin of the OIR emission in the hard state if the hard X-rays are the source of irradiation, and suggest that the onset of the compact jets emission starts by the low frequencies.

\end{enumerate} 

We finally think that further investigations with good enough multi-wavelenght data to follow the evolution of the break frequency together with broadband SED fitting will allow us to disentangle the dominant emission processes in the various energy bands.

\section*{Acknowledgments}

MC warmly thanks Jerome Rodriguez for many useful comments and discussions, Guillaume Dubus for his precious help on accretion disc theory, Mike Nowak for useful discussions on Galactic ridge emission
and Sera Markoff for useful discussions. The authors also thank the anonymous referee for constructive comments. 

MMB and CDB would like to acknowledge support from NSF grant AST-0707627. JAT would like to acknowledge partial support from NASA ADP grant NNX08AJ59G. This research has made use of data obtained through the High Energy Astrophysics Science Archive Research Center (HEASARC), provided by NASA's Goddard Space Flight Center. The ATCA is part of the Australia Telescope funded by the Commonwealth of Australia for operation as a National Facility managed by CSIRO.

\bibliographystyle{mn2e}
\bibliography{biblio}

\bsp

\label{lastpage}

\end{document}